\documentstyle[12pt]{article}
\newcommand{\be}{\begin{equation}}
\newcommand{\ee}{\end{equation}}
\newcommand{\ba}{\begin{eqnarray}}
\newcommand{\ea}{\end{eqnarray}}

\date{}

\begin{document}
\begin{flushright}
Preprint SPbU-IP-94-05
\end{flushright}

\vspace{1.5cm}

\begin{center}
{\Large\bf Polynomial SUSY in Quantum
Mechanics and Second Derivative Darboux Transformation}\\
\vspace{1.cm}
{\large\bf A. A. Andrianov, M.V.Ioffe and
D.N.Nishnianidze} \footnote{On leave of absence from
Kutaisi Polytechnic University, Georgia}\\
\vspace{1.cm}
Department of Theoretical Physics,
University of Sankt-Petersburg,198904 Sankt-Petersburg, Russia.
\footnote{E-mail:  ANDRIANOV1@TFLD1.SAMSON.SPB.SU
and IOFFE@TFLD1.SAMSON.SPB.SU}
\end{center}

\vspace{2cm}

{\small  We give the classification of second-order polynomial
SUSY Quantum Mechanics in one and two dimensions. The
particular attention is paid to the irreducible supercharges
which cannot be built by repetition  of ordinary Darboux
transformations. In two dimensions it is found that the
binomial superalgebra leads to the dynamic symmetry generated
by a central charge operator.}

\newpage
\vspace{1.cm}

\section{Introduction}
\vspace{.5cm}
\hspace*{3ex}The Supersymmetrical Quantum Mechanics (SSQM)
\cite{Gend},\cite{Lahiri} represents a concise algebraic form
of spectral equivalence between different hamiltonian quantum
systems realized by means of Darboux transformation
\cite{Darboux},\cite{Infeld}. Certainly SSQM is used also for
verification of some properties predicted by SUSY in the QFT
\cite{Witten}.

   On the other hand the variety of extensions of SUSY-algebra
seems to be broader in QM than in the QFT. Recently the
polynomial SUSY-algebra has been proposed and elaborated in the
one-dimensional SSQM \cite{AIS}. Its consequences for the
scattering characteristics of spectrally-equivalent systems
have been outlined in \cite{ACDI}.

   The aim of our paper is to give the complete classification
of second-order polynomial SSQM in one-dimensional and to
generalize the polynomial SUSY onto two-dimensional QM. After
short description of related superalgebra we select out the
subclass of reducible SUSY algebras which is built by
repetition of two single Darboux transformations or
equivalently by gluing of two ordinary SUSY systems \cite{AIS},
(see also \cite{baye}). We distinguish the irreducible subclass
which cannot be factorized in this way. In the one-dimensional
case both subclasses are rich of representatives as it is shown
in Sect.2. The two-dimensional version of the polynomial SUSY
is built in Sect.3 and it is found that the reducible subclass
is restricted with potentials combining the harmonic oscillator
and the centrifugal-type terms.

    The irreducible class appears to be more significant and is
generated by the second derivative supercharges. The set of
nonlinear equations on potentials and coefficient functions of
supercharges are derived. In Sect.4 the corresponding binomial
superalgebra is analyzed and nontrivial operator of dynamic
central charge is discovered. When the senior derivatives of
supercharge form the Laplacian the separation of variables is
provided in particular coordinate systems. The explicit form of
potentials is described and the operators of supercharge and
central charge are evaluated. In Sect.5 the second derivative
supercharges with metric $g_{ik}({\vec x})$ are introduced and
their possible structure is analyzed. It is found that in the
general case the polynomial superalgebra defines the dynamic
symmetry operator of Hamiltonian system. Therefore the building
of isospectral systems is tightly connected with another
problem namely with a search for dynamic symmetry operator~$R$.

\section{Polynomial SUSY in one dimension }
\vspace{.5cm}
\hspace*{3ex}Let us remind the basic notations of the ordinary
 SUSY QM \cite{Gend},\cite{Witten} and describe its polynomial
extension \cite{AIS}. The intertwining relations between two
Hamiltonians with equivalent spectra are realized by means of
well-known Darboux transformation \cite{Darboux}:
\ba
h^{(1)} = q^+q^- = -\partial^2 + V^{(1)}(x);&\quad &
h^{(1)} \Psi_n^{(1)}(x) = E_n \Psi_n^{(1)}(x);\nonumber\\
h^{(2)} = q^-q^+ = -\partial^2 + V^{(2)}(x);&&
h^{(2)} \Psi_n^{(2)}(x) = E_n \Psi_n^{(2)}(x);\nonumber\\
h^{(1)}q^+ = q^+h^{(2)};&&
q^-h^{(1)} = h^{(2)}q^-;\label{stSUSY}\\
q^+ = -\partial + \partial\chi(x);&&
q^- = (q^+)^{\dagger} = \partial + \partial\chi(x);\nonumber\\
\Psi_n^{(2)}(x) = q^- \Psi_n^{(1)}(x);&&
\Psi_n^{(1)}(x) = q^+ \Psi_n^{(2)}(x),\nonumber
\ea
where $\partial \equiv d/dx.$

The concise algebraic form of spectral equivalence is given
by superalgebra for the partners $h^{(1)}, h^{(2)}$ and
off-diagonal supercharges:
\ba
 H =
\left( \begin{array}{cc}h^{(1)}&0\\
0& h^{(2)}
\end{array} \right); \qquad
 Q^+ =(Q^-)^{\dagger} = \left( \begin{array}{cc}
                        0&0\\
                        q^-&0
                        \end{array} \right);\nonumber\\
  \{Q^+,Q^-\} = H;\quad (Q^+)^2 =(Q^-)^2 = 0;\quad [H ,Q^{\pm}]
  =0.
                                         \label{strepr}
\ea

The superpotential $\chi(x)$ is defined by an arbitrary solution
$\Psi(x)$ of the Schr\"odinger equation  $h^{(1)}
\Psi(x) = E \Psi(x)$
\be
\chi(x) \equiv - \ln \Psi(x).
\ee
If this solution is nodeless one has almost coinciding spectra
of $h^{(1)}$ and $h^{(2)}$ or, equivalently, double degeneracy
of the energy spectrum of $H$. The polynomial superalgebra is
created \cite{AIS} by the intertwining operators of higher
order in derivatives. Let as consider the second-derivative
superalgebra generated by the following operators
\be
q^+ = \partial^2 - 2f(x) \partial + b(x);\quad q^- =
(q^+)^{\dagger}.
\ee
The conserving supercharges $[H, Q^{\pm}] = 0$
determine the Hamiltonian as follows :
\be
\{Q^+, Q^-\} = (H + a)^2 + d,    \label{2-deralgebra}
\ee
where the potentials can be expressed in terms
of a real coefficient function $f(x)$ :
\be
 V^{(1,2)} = \mp2f'(x) + f(x)^2 + \frac{f''(x)}{2f(x)} -
 \biggl(\frac{f'(x)}{2f(x)}\biggr)^2 - \frac{d}{4f(x)^2} - a.
\label{V}
\ee
  The intertwining relations for $H$ require that
\be
b(x) = - f'(x) + f(x)^2 - \frac{f''(x)}{2f(x)} +
 \biggl(\frac{f'(x)}{2f(x)}\biggr)^2 + \frac{d}{4f(x)^2}.
         \label{b}
\ee

Depending on the sign of $d$ one finds \cite{ACDI} either the
reducible algebra $(d<0)$ or the irreducible one $(d>0)$. In
the reducible case there exists an intermediate Hamiltonian
$h$ which is a superpartner of both $h^{(1)}$ and $h^{(2)}$
with the ordinary superalgebra, respectively the second-order
Darboux transformation can be factorized into a product of two
ordinary Darboux transformation (see details in
\cite{AIS},\cite{ACDI}):
\ba
&& q^+ = q_1^+ q_2^+ = (- \partial + W_1 )(- \partial + W_2
);\nonumber \\
&& W_1(x) \equiv \partial \chi_1(x); \quad
   W_2(x) \equiv \partial \chi_2(x);\label{redcase} \\
&&h^{(1)} = q_1^+q_1^- + c/2; \quad h = q_1^- q_1^+ +c/2 = q_2^+
q_2^- - c/2; \quad h^{(2)} = q_2^- q_2^+ - c/2;  \nonumber\\
&& W_{1,2}= \pm \frac{2f'(x) -c}{4f(x)} - f(x).\nonumber
\ea
which corresponds to $a=0,\quad c^2=-4d$.
In the irreducible case $(d > 0)$ $c$ becomes imaginary and
the hermitian intermediate Hamiltonian $h$ does not exist. In
fact the analytic continuation from $c$ to $ic$ leads to a
complex potential for $h$. Respectively $W_1$ and $W_2$ are
complex functions and $q_i^{\pm} = \mp \partial + W_i\quad
(i=1,2)$, but $q_i^+ \not = (q_i^-)^\dagger $. Nevertheless the
second-order supercharges are hermitian
\be
q_1^+q_2^+ = (q_2^-q_1^-)^\dagger. \label{herq}
\ee
{}From the hermiticity of $h^{(1)},h^{(2)}$ and
Eqs.(\ref{redcase}),(\ref{herq}) it follows that
 $$Re W_1 =-\frac{f'(x)}{2f(x)} + f(x);\quad Im W_1
 =-\frac{c}{4f(x)},$$
where $f(x)$ is a real function as before and potentials are
described by Eq.(\ref{V}).

In general there is the large variety of pairs of
potentials $V^{(1)}$ and $V^{(2)}$  which obey Eq.(\ref{V}) for
any sign of $d$. The higher-order polynomial SUSY algebra can
be created by composition of ordinary (first-order in
derivatives) SUSY transformations and of irreducible ones,
which are of second-order in derivatives:  \be \{ Q^+,Q^- \} =
 \prod_{i+2j=n}(H + c_j )((H + a_i)^2 + d_j);\quad d_j > 0.
\ee

\section {Two-dimensional Darboux transformations of
second-order in derivatives}
\vspace{.5cm}
\hspace*{3ex} The conventional two-dimensional SUSY algebra
provides the equivalence of energy spectra between a pair of
two scalar Hamiltonians $h^{(1)}$, $\widetilde h^{(1)}$ and $2 \times2$
matrix Hamiltonian $h_{ik}$ $(i,k = 1,2)$ \cite{ABI}:
\begin{eqnarray*}
h^{(1)}&=&-\partial_l^2 + V^{(1)}({\bf x}) = -\partial_l^2 +
(\partial_l\chi)^2 - \partial_l^2\chi  ;\\
\widetilde h^{(1)}&=&-\partial_l^2 +\widetilde V^{(1)}({\bf x}) = -\partial_l^2
+ (\partial_l\chi)^2 + \partial_l^2\chi  ;\\ h_{ik}&=&
-\delta_{ik} \partial_l^2 + \delta_{ik}((\partial_l\chi)^2 -
 \partial_l^2\chi) + 2\partial_i\partial_k\chi,
 \end{eqnarray*}
where the summation on index $l$ is implied, $\partial_l^2
\equiv \partial_1^2 + \partial_2^2$.

 The superalgebra is represented by intertwining relations:
\begin{eqnarray*}
 h^{(1)}q_i^+&=&q_k^+h_{ki};\quad h_{ik}q_k^- =
 q_i^-h^{(1)}; \quad q_i^{\pm} \equiv \mp \partial_i +
 \partial_i \chi (\vec x);\\ h_{ik}p_k^-&=&p_i^-\widetilde
 h^{(1)} ;\quad p_k^+h_{ki} = \widetilde h^{(1)}p_i^+,\\
q_i^{\pm}p_i^{\mp}&=&0 ;\quad
p_i^-=(p_i^+)^{\dagger}=\epsilon_{ik}q_k^+;\quad
\epsilon_{ik}=-\epsilon_{ki}.
\end{eqnarray*}

 The energy spectra of $h^{(1)}$ and $\widetilde h^{(1)}$ are different
except for particular cases \cite{Pauli},\cite{Chun}. However
it is of interest to extend this superalgebra so that the
spectral equivalence could be established between some scalar
hamiltonians.  One can attempt to build the polynomial
superalgebra similarly to the one-dimensional reducible scheme
:
\begin{displaymath}
\left( \begin{array}{cc}
h^{(1)}&0\\
0&\widetilde h^{(1)}
\end{array} \right) \longrightarrow h_{ik} = h_{ik}^{(*)} - 4cI
\longrightarrow
\left( \begin{array}{cc}
h^{(2)}&0\\
0&\widetilde h^{(2)}
\end{array} \right).
\end{displaymath}

In result we arrive to the following intertwining relations
between upper components $h^{(1)}$ and $h^{(2)}$
\be
h^{(1)}q^+ = q^+h^{(2)};  \quad q^-h^{(1)} = h^{(2)}q^-,
\label{intertw}
\ee
where
\ba
h^{(2)} = \widetilde q_i^+\widetilde q_i^- - 4cI;\quad
q^+ = q_i^+\widetilde q_i^-= (-\partial_i +
\partial_i\chi)(\partial_i + \partial_i\widetilde\chi);\quad q^- =
\widetilde q_i^+q_i^-.
\ea

   We have investigated the consistency
conditions for superpotentials and have found the following
solutions in this reducible case
\ba
&&\chi = \frac{1}{2}a\rho^2 - \frac{c}{a}ln\rho +
\xi(\varphi);\nonumber\\
&&\widetilde{\chi} =
-\frac{1}{2}a\rho^2 - \frac{c}{a}ln\rho + \xi(\varphi) + b;
\nonumber\\ &&V^{(1)} = a^2\rho^2 +
\frac{1}{\rho^2}\Biggl[\xi'^2 - \xi'' + \frac{c^2}{a^2}\Biggr]
- 2(a + c);\nonumber\\ &&V^{(2)} = a^2\rho^2 +
\frac{1}{\rho^2}\Biggl[\xi'^2 - \xi'' + \frac{c^2}{a^2}\Biggr]
+ 2(a - c),\nonumber
\ea
where $a, b$ are constants and
$\xi(\varphi)$ is arbitrary function of angular variable
$\varphi$.

   In two dimensions the irreducible class of
second-derivative superalgebra with intertwining components
\ba
q^+ = -\triangle + C_i \partial_i + B ; \quad  q^- =
(q^+)^{\dagger}; \quad \triangle \equiv \partial_l^2
\label{secch}
\ea
can be constructed when imposing the
conditions (\ref{intertw}) irrespectively of the existence of
superpartners $\widetilde h^{(1)}$  and $\widetilde h^{(2)}$.
Thus we do not assume that  $q^+ = q_i^+\widetilde q_i^-$ where
  $[q_i^+,q_j^+]=[\widetilde q_i^+,\widetilde q_j^+]=0$ (the
integrability condition). At this point the two-dimensional
case is drastically different from the one-dimensional case.

   The intertwining between $h^{(1)}$ and $h^{(2)}$ leads to six
equations on five real functions $(C_{1,2},B,V^{(1,2)})$ :
\ba
&&-\partial_{\mu}C_{\nu} - \partial_{\nu}C_{\mu} +
(V^{(2}-V^{(1)})\delta_{\mu\nu} = 0; \nonumber\\
&&\triangle C_{\nu} - 2\partial_{\nu}(V^{(2)}-B) -
(V^{(1)}-V^{(2)})C_{\nu} = 0; \label{2int}\\
&&-\triangle (V^{(2)}-B) - (V^{(1)}-V^{(2)})B +
C_{\nu}\partial_{\nu}V^{(2)} = 0.\nonumber
\ea

Certainly they cannot be satisfied for arbitrary potentials
$V^{(1)}$ or $V^{(2)}$.  Indeed the solutions can be found
explicitly and read :
\ba
&&C^2 \equiv (C_1+iC_2)^2 = \alpha z^2 + 8\beta z + \gamma;
\quad z\equiv x_1 + i x_2;\nonumber\\
&&V^{(2)}-B = \frac{1}{4}\alpha\mid z\mid^2 + z\bar \beta +
\bar z\beta + \frac{1}{4}\mid C\mid^2 - \eta \label{2dim};\\
&&V^{(2)}-V^{(1)} = \partial_zC + \partial_{\bar z}\bar C,
\nonumber
\ea
where $\alpha$, $\eta$ are real constants and
$\beta$, $\gamma$  are complex constants. The further
evaluation of solutions for $V^{(1,2)}$ is based on the
equation :
\be
(C\partial_z + \bar C\partial_{\bar z})(B\mid
C\mid^2) = G\mid C\mid^2,
\ee
where
\be G = \alpha +
(\partial_zC)(\partial_{\bar z}\bar C) - \frac{\alpha}{2}(\bar
z C + z\bar C) - 2(\bar \beta C + \beta \bar C ).
\ee

In terms of variables
$$
\tau_1 = \int\frac{dz}{C} + \int\frac{d\bar z}{\bar C}; \quad
i\tau_2 = \int\frac{dz}{C} - \int\frac{d\bar z}{\bar C},
$$
one obtains
\be
B = \frac{1}{2\mid C\mid^2}\int G\mid C\mid^2d\tau_1 +
\frac{F(\tau_2)}{\mid C\mid^2}, \label{B}
\ee
where $F(\tau_2)$ is an arbitrary function. Thus we find that
the irreducible second-derivative SUSY algebra is realized in
the class of potentials which is much
broader comparing to the reducible case.
\section{The second-derivative superalgebra}
\vspace{.5cm}
\hspace*{3ex}Let us describe superalgebra relations for
the irreducible Darboux transformations (\ref{intertw}). Since
supercharges $Q^{\pm}$(with components (\ref{secch})) commute
with the Hamiltonian $H$ one expects that the closing of
superalgebra leads to the symmetry operator $R$ (central
charge) :  $$ \{Q^+,Q^-\} = F(H,R);\qquad [H,R]=0.$$

For the supercharges (\ref{secch}) one finds that
\be
F(H,R) = H^2 + R + 2\eta H, \label{polyn}
\ee
where $R$ is a diagonal operator:
$$
R = \left( \begin{array}{cc}
       R_1 & 0 \\
       0   & R_2 \\
      \end{array} \right).
$$
Its components
\ba
&&R_1 = 2\bigl(\alpha \mid z\mid^2 + 4(\bar \beta z + \beta
\bar z )\bigr)\partial_z \partial_{\bar z} - C^2\partial_z^2 -
\bar C ^2 \partial_{\bar z}^2 -\nonumber\\
&&C (\partial_z C)\partial_z -
\bar C (\partial_{\bar z} \bar C)\partial_{\bar z} + (C\partial_z +
\bar C\partial_{\bar z})(B + V^{(1)}) +\nonumber\\
&&B^2 - V^{(1)2} -2\eta V^{(1)}; \label{symm1}
\ea
\ba
&&R_2 = 2\bigl(\alpha \mid z\mid^2 +
4(\bar \beta z + \beta \bar z )\bigr)\partial_z \partial_{\bar
z} - C^2\partial_z^2 - \bar C ^2\partial_{\bar z}^2
-\nonumber\\ &&C(\partial_z C)\partial_z - \bar
C (\partial_{\bar z} \bar C )\partial_{\bar z} - (C\partial_z +
\bar C \partial_{\bar z})(B - V^{(2)}) +\nonumber\\
&&B^2 - V^{(2)2} -2\eta V^{(2)} \label{symm2}
\ea
 are the symmetry operators of Hamiltonians
$h^{(1)}$, $h^{(2)}$ respectively.

 Let us derive the explicit solution of Eqs.(\ref{2int}) and
the related operators $H, Q^{\pm}, R$ . They have different
form depending on values of $\alpha$ and $\beta$  in
Eqs.(\ref{2dim}).

i) If  $\alpha=0$, $\beta\not=0$, then  $\gamma$ can be
eliminated by translation $z\rightarrow
z - (\gamma/8\beta)$. In this case the variables
\ba
\tau_1 = \sqrt{\frac{z}{2\beta}} + \sqrt{\frac{\bar
z}{2\bar \beta}}; \quad \tau_2 =
i\Bigl(\sqrt{\frac{\bar z}{2\bar \beta}} -
\sqrt{\frac{z}{2\beta}}\Bigr)
\ea
can be referred to the
conventional parabolic coordinates \cite{Miller}.  From the
Eq.(\ref{B}) one obtains that
\be
B =\frac{2\tau_1 - \mid
\beta\mid^2\tau_1^4 - 2\mid \beta \mid^2 \tau_1^2 \tau_2^2 +
F(\tau_2)}{\tau_1^2 + \tau_2^2}
\ee
and from Eqs.(\ref{2dim})
one finds the following expressions for potentials
\ba
V^{(1)}= \frac{-2\tau_1 + \mid \beta \mid^2 \tau_1^4 +
F(\tau_2)}{\tau_1^2 + \tau_2^2} - \eta; \label{V1}\\ V^{(2)}
=\frac{2\tau_1 + \mid \beta \mid^2 \tau_1^4 +
F(\tau_2)}{\tau_1^2 + \tau_2^2} - \eta. \label{V2}
\ea
In terms of variables  $\tau_1$, $\tau_2$ the Laplacian is
separable
 $$ \triangle = \frac{1}{\mid \beta \mid^2(\tau_1^2 +
\tau_2^2)}(\partial_{\tau_1}^2 + \partial_{\tau_2}^2)$$
with the same factor as in Eqs.(\ref{V1}),(\ref{V2}). Hence the
spectral problem for both Hamiltonians $h\Psi=E\Psi$ can be
solved by $R$-separation \cite{Miller} of variables (further on
we omit the indexes of $h, R , \Psi$ for brevity). Namely one
can decompose the corresponding eigenfunctions for a given $E$
into a sum
\be
 \Psi = \sum_n{\nu_n \phi_{1n}(\tau_1)
\phi_{2n}(\tau_2)}, \label{wave}
\ee
 where $\nu_n$ are constants and $\phi_{1n}(\tau_1)$,
$\phi_{2n}(\tau_2)$ are solutions of one-dimensional equations :
\ba
&& -\phi_{1n}''(\tau_1) + \mid \beta \mid^2[-(E+\eta)\tau_1^2
\mp 2\tau_1 + \mid\beta\mid^2\tau_1^4] \phi_{1n}(\tau_1) =
\frac{\lambda_n}{4}\phi_{1n}(\tau_1); \label{phi1}\\
&&-\phi_{2n}''(\tau_2) + \mid\beta\mid^2[-(E+\eta)\tau_2^2 +
F(\tau_2)] \phi_{2n} =
-\frac{\lambda_n}{4}\phi_{2n}(\tau_2).
\ea
where the upper (lower) sign in Eq.(\ref{phi1}) corresponds to
$h^{(1)}(h^{(2)}$ and $\lambda_n$ is a separation constant
which serves as a spectral parameter for the symmetry operator
$R$ :  $R\phi_{1n}(\tau_1)\phi_{2n}(\tau_2) = (\lambda_n -
\eta^2)\phi_{1n}(\tau_1)\phi_{2n}(\tau_2)$. Depending on
properties of the function $F(\tau_2)$ one can obtain either the
finite-dimensional subspace $\{\lambda_i\}$ of degeneracy for
the energy spectrum or even the infinite-dimensional one.
However the spectral values of $h$
and $R$ form the complete set of quantum numbers (the integral
of motion) which characterize uniquely the quantum state.

 ii) Let us describe the second case  $\beta=0$, $\alpha > 0$
when the appropriate coordinates are elliptic ones:
\ba
&&\tau_1 = \frac{1}{\sqrt\alpha}ln\biggl(z + \sqrt{z^2 +
\frac{\gamma}{\alpha}}\biggr)\biggl(\bar z + \sqrt{\bar z ^2 +
\frac{\bar \gamma}{\alpha}}\biggr);  \label{tau1}\\
&&\tau_2 = - \frac{i}{\sqrt\alpha}ln\frac{z + \sqrt{z^2 +
\frac{\gamma}{\alpha}}}{\bar z + \sqrt{\bar z ^2 +
\frac{\bar \gamma}{\alpha}}}.       \label{tau2}
\ea
In a full analogy to case i) one can find
\ba
&&B = \frac{1}{2(f_1 + f_2)}(2\partial_{\tau_1}f_1 -
\frac{1}{2}f_1^2 - f_1f_2 + F(\tau_2)); \label{spt1} \\
&&V^{(1)} = \frac{1}{2(f_1 + f_2)}(-2\partial_{\tau_1}f_1 +
\frac{1}{2}f_1^2 + F(\tau_2)) - \eta; \label{spt2} \\
&&V^{(2)} = \frac{1}{2(f_1 + f_2)}(2\partial_{\tau_1}f_1 +
\frac{1}{2}f_1^2 + F(\tau_2)) - \eta ,   \label{spt3}
\ea
where
\ba
&&f_1 = \frac{1}{4}\biggl(\alpha\cdot exp(\sqrt{\alpha}\tau_1)
+ \frac{\mid\gamma\mid^2}{\alpha} exp(-\sqrt{\alpha}\tau_1)
\biggr); \label{ef1}\\
&&f_2 = \frac{1}{4}\biggl(\bar \gamma
\cdot exp(i\sqrt{\alpha}\tau_2) + \gamma\cdot
exp(-i\sqrt{\alpha}\tau_2)\biggr).   \label{ef2}
\ea
This coordinates allow to
separate variables as well that leads to the solution for the
wave function in the form (\ref{wave}) where now
\ba
-\phi_{1n}''(\tau_1) +
\frac{1}{4}\Biggl(\mp \partial_{\tau_1}f_1 + \frac{1}{2}f_1^2 -
 (E + \eta)f_1\Biggr)\phi_{1n}(\tau_1) =
 \frac{\lambda_n}{4}\phi_{1n}(\tau_1); \label{phi2}\\
-\phi_{2n}''(\tau_2) + \frac{1}{4}\Biggl(\frac{F(\tau_2)}{2} -
 (E + \eta)f_2\Biggr)\phi_{2n}(\tau_2) =
-\frac{\lambda_n}{4}\phi_{2n}(\tau_2)
\ea
and $\lambda_n$ is again eigenvalue of
the symmetry operator Eqs.(\ref{symm1}),(\ref{symm2}).

The analysis of the case $\beta = 0,\quad \alpha < 0$ is similar
and its result can be formulated as follows. One has to replace
real $(\tau_1, \tau_2)$ by imaginary $(- i\tau_2, -i\tau_1)$ in
Eqs.(\ref{tau1}), (\ref{tau2}) and respectively
$\left( f_1(\tau_1), f_2(\tau_2)\right)$ by
$- f_2(\tau_2), -f_1(\tau_1)$ in Eqs.(\ref{ef1}), (\ref{ef2}).
Then the relations for potentials $B, V_1, V_2$ are given again
by Eqs.(\ref{spt1}) - (\ref{spt3}).

We remark that the supercharge is factorizable into a
composition of two ordinary SS-operators  $(q_i^+\widetilde q_i^-
)$ for the special choice $\alpha>0$,
$\beta=\gamma=0$ that corresponds to the separation of
variables in the polar coordinates (see Sect.3). We conclude
that from the supersymmetry or equivalently from the
intertwining of two Hamiltonians one inevitably recovers the
hidden dynamical symmetry realized by $R$ and furthermore the
$R$-separation of variables.

\section{General case and discussion}
\vspace{.5cm}
\hspace*{3ex}Let us describe the natural extension of the
second derivative Darboux transformation which is generated by
the operator with metric
\be
q^+ = g_{ik}{(\vec x)}\partial_i \partial_k + \widetilde
C_i\partial_i + \widetilde B. \label{gench}
\ee
The intertwining relations (\ref{intertw}) determined
completely the form of metric  $g_{ik}{(\vec x)}$ which
satisfies the following equation:
$$ \partial_l g_{ik}{(\vec
x)} + \partial_i g_{lk}{(\vec x)} + \partial_k g_{il}{(\vec x)}
= 0. $$ Its solution can be easily found
\ba
&&g_{11} =\widetilde{\alpha}y^2 + \widetilde a_1 y +
\widetilde b_1; \nonumber\\ &&g_{22} = \widetilde{\alpha}x^2 +
\widetilde a_2 x + \widetilde b_2; \nonumber\\ &&g_{22}
=-\frac{1}{2}(2\widetilde{\alpha}x y + \widetilde a_1 x +
\widetilde a_2 y) + \widetilde b_3. \nonumber
\ea

Thus one can see that in senior derivatives the possible
supercharges belong to the $E(2)$-universal enveloping
algebra \cite{Miller} in which, at the level of second order, we
distinguish three different possibilities :
\ba
&&q^{(1)+} = \alpha P_1^2 + \gamma \triangle + \widetilde C_i
\partial_i + \widetilde B; \nonumber\\
&&q^{(2)+} = \alpha \{J,P_1\} + \gamma \triangle + \widetilde C_i
\partial_i + \widetilde B; \nonumber\\
&&q^{(3)+} = \alpha J^2 + \beta P_1^2 + \gamma \triangle +
\widetilde C_i \partial_i + \widetilde B, \nonumber
\ea
where $J$ and $ \vec P $ are rotation and translation
generators,respectively,and $ \alpha\not=0$.

The coefficients of supercharge Eq.(\ref{gench}) and the
potentials $V^{(1),(2)}$ obey the modified equations:
\ba
&&\partial_i{\widetilde C_k} + \partial_k{\widetilde C_i} + \triangle
g_{ik} - (V^{(1)} - V^{(2)})g_{ik} = 0; \nonumber\\
&&\triangle{\widetilde C_i} + 2\partial_i\widetilde B + 2
g_{ik}\partial_k V^{(2)} - (V^{(1)} - V^{(2)})\widetilde
C_i=0;\nonumber\\
&&\triangle{\widetilde B} + g_{ik}\partial_k\partial_i V^{(2)} +
\widetilde C_i\partial_i V^{(2)} - (V^{(1)} - V^{(2)})\widetilde B =
0.\nonumber
\ea
The generalized superalgebra evidently yields
to the symmetry operator for the Hamiltonian $H$:
$$\{Q^+,Q^-\} = \widetilde R;  \qquad [\widetilde R,H] = 0. $$
However this operator creates the dynamical symmetry of higher
order which cannot be in general represented by a polynomial of
the Hamiltonian and of second order symmetry operator (similar
to Eq.(\ref{polyn})). But in any case the closing of SUSY
algebra leads to the $R$-separation of variables in the
spectral problem for the Hamiltonian $H$ (to the integrability
of the corresponding dynamical system).

This work was supported partially by the Russian Foundation for
Fundamental Research. One of us (M.V.I.) is indebted to the
International Science Foundation (G.Soros Foundation) and to
the American Physical Society for the financial support. One of
us (D.N.N.) is grateful to Prof. A.Kostava for encouragement
and support.

\end{document}